\documentclass[prb,twocolumn,superscriptaddress,showpacs,superbib]{revtex4}
\usepackage{graphicx, textcomp}
\usepackage{amssymb}

\usepackage{neel}%
\usepackage{units}%

\usepackage{hyperref}
\definecolor{link_green}{rgb}{0.0,0.7,0.0}
\hypersetup{pagebackref,breaklinks,citecolor=link_green,colorlinks,bookmarks,bookmarksopen,
  pdfauthor="Olivier Fruchart",pdfstartview=FitH}
\usepackage[all]{hypcap}

\graphicspath{{fig/}}

\begin{document}

\title{Growth mode, magnetic and magneto-optical properties of pulsed-laser-deposited
Au/Co/Au(111) trilayers}

\author{C. Clavero}

\email[Current address: Department of Applied Science, College of William \& Mary, Williamsburg, Virginia 23187, USA.]{cclavero@wm.edu}

\affiliation{Instituto de Microelectr\'{o}nica de Madrid-IMM (CNM-CSIC), Isaac
Newton 8 (PTM), 28760 Tres Cantos, Madrid, Spain}

\author{A. Cebollada}

\affiliation{Instituto de Microelectr\'{o}nica de Madrid-IMM (CNM-CSIC), Isaac
Newton 8 (PTM), 28760 Tres Cantos, Madrid, Spain}

\author{G. Armelles}

\affiliation{Instituto de Microelectr\'{o}nica de Madrid-IMM (CNM-CSIC), Isaac
Newton 8 (PTM), 28760 Tres Cantos, Madrid, Spain}

\author{O. Fruchart}

\affiliation{Institut N\'{e}el (CNRS \& UJF), BP166, F-38042 Grenoble, Cedex 9, France}

\date{\today}

\begin{abstract}

The growth mode, magnetic and magneto-optical properties of epitaxial Au/Co/Au(111) ultrathin trilayers grown by pulsed-laser deposition (PLD) under ultra-high
vacuum are presented. Sapphire wafers buffered with a single-crystalline Mo(110) bilayer were used as substrates. Owing to PLD-induced interfacial intermixing at the lower Co/Au(111) interface, a layer-by-layer growth mode is promoted. Surprisingly, despite this intermixing, ferromagnetic behavior is found at room temperature for coverings starting at 1 atomic layer (AL). The films display perpendicular magnetization with anisotropy constants reduced by 50\% compared to
TD-grown or electrodeposited films, and with a coercivity more than one order of magnitude
lower ($\lesssim$ 5 mT). The magneto-optical (MO) response in the low Co thickness range is dominated by Au/Co interface contributions. For thicknesses starting at 3 AL Co, the MO response has a linear dependence with the Co thickness, indicative of a continuous-film-like MO behavior.

\end{abstract}


\maketitle


\section{Introduction}

Owing to its ability to essentially preserve the stoichiometry of targets during evaporation,
Pulsed-Laser Deposition~(PLD) is often used for the growth of materials with a complex
composition, particularly oxides such as high-Tc superconductors, ferroelectrics and manganites\cite{Horwitz1998}.
When performed under ultra-high vacuum conditions PLD is also suitable for the epitaxy of
metals\cite{bib-SHE2004}, although it is seldom used for that purpose. There are two main differences between PLD and Thermal Deposition (TD) concerning the epitaxy of metals. The first difference is the possibility to force
layer-by-layer growth in some cases\cite{bib-OHR1999}, which is a positive aspect. This fact arises from the several orders of magnitude higher instantaneous deposition rate during the laser pulse duration and from the increased kinetic energy of the ejected species (up to several eV) as compared with TD\cite{bib-SIN1990}. The second aspect is to potentially induce some intermixing at interfaces\cite{bib-MEY2007}, which may be a drawback to reach certain physical properties depending on the smoothness of interfaces. Evidence of an increased tendency of intermixing at interfaces in structures grown by PLD versus TD has been given previously by X-ray diffraction\cite{bib-MEY2007}, and it is generally thought to result from the energy carried by the atoms or ions evaporated from the target and heated upon further interaction with the laser in the plume\cite{bib-SIN1990}. Here we report on the growth by PLD and the resulting magnetic and magneto-optical (MO) properties of epitaxial Au/Co/Au trilayers, motivated by the observed perpendicular magnetic anisotropy in TD grown\cite{bib-CHA1986} or electrodeposition(ED)\cite{bib-CAG2001} Co/Au(111) films capped with various materials.

In this work the Au surface used as substrate is a thin (111)
film deposited on a buffer layer of refractory metal
[Mo(110)] epitaxially grown on Sapphire $(11\overline20)$. For this
particular set and order of elements (\emph{i.e.}, Co deposited on
Au) we evidence by scanning tunneling microscopy some
intermixing at the Au/Co interface, inducing a layer-by-
layer growth for Co as compared to a two-atomic-layer-
high-island growth for TD. Interestingly, despite the intermixing ferromagnetic behavior is found at room temperature for coverings starting at 1 atomic layer (AL). An easy
axis of magnetization is found perpendicular to the plane for Co coverings between 1 and 5 AL,
with magnetic anisotropy constants similar to their TD
or ED counterparts, however with a much lower coercivity. The spectroscopic magneto-optical activity of the films has also been measured and modeled.

\section{Experimental}

The samples were grown by pulsed laser deposition (PLD) under ultrahigh vacuum (UHV) inside a
three-chamber setup\cite{bib-FRU2007}. The first chamber is devoted to preparation and analysis. It is equipped with a heater for sample degassing up to 800 $^\circ$C, a sputtering gun and an Auger electron
analyzer. The base pressure is $2\times 10^{-10}$ Torr. The second chamber is the deposition
chamber. It is equipped with a \unit[10]{keV} reflection high-energy electron diffraction (RHEED)
setup, a quartz microbalance, and a sample heating similar to that of the first chamber. The base
pressure is $2-3\times 10^{-11}$ Torr and in the $10^{-10}$ Torr range during laser deposition. A third chamber is dedicated to Scanning Tunneling Microscopy (STM) with a room temperature (RT)
Omicron-1 setup (base pressure $5\times 10^{-11}$ Torr). A Nd-YAG laser with a \unit[10]{ns} pulse duration and a \unit[10]{Hz} frequency was used. The targets are first mechanically polished \exsitu and then surface-cleaned \insitu through laser ablation until no gas contaminant is found on the target, as controlled by Auger spectroscopy. The targets are cleaned at the same fluence as that used during the growth~(about $\unit[1]{J/cm^2}$). This sequence is chosen for each element just above the evaporation threshold, so as both to avoid the formation
of droplets\cite{bib-CHE1988} and to minimize the energy carried by evaporated individual atoms or ions\cite{bib-SIN1990}. Under these conditions the typical growth rate on the sample is 1\AA/s. More details can be found in Ref.\cite{bib-FRU2007}.

Hysteresis loops were carried out by means of a Superconducting Quantum Interference Device
(SQUID) magnetometer at low and RT. The magneto-optical (MO)
polar response of the samples was studied experimentally in the spectral range from 1.4 to
$\unit[4.3]{eV}$, using a spectrometer described elsewhere\cite{bib-CLA2007}.

\section{Au(111) surface preparation}

We used commercial sapphire$(11\overline20)$ wafers with a miscut angle smaller than $0.1\,^{\circ}$. A
$\thicknm{10}$-thick buffer layer of Mo$(110)$ was first deposited following an optimized
procedure\cite{bib-FRU2007,bib-FRU1998b}. Its surface is
single-crystalline and displays $\thicknm{\approx200}$-wide terraces separated by monoatomic
steps. The steps separation and orientation are determined by the miscut of Sapphire, which is uniform on
a two-inch wafer, however varies from one wafer to another. This residual miscut, on the average smaller than
that typically found on metal single-crystals, does not influence the growth mode and thus presumably neither
the magnetic properties.

\begin{figure}
\includegraphics[width=3.5in]{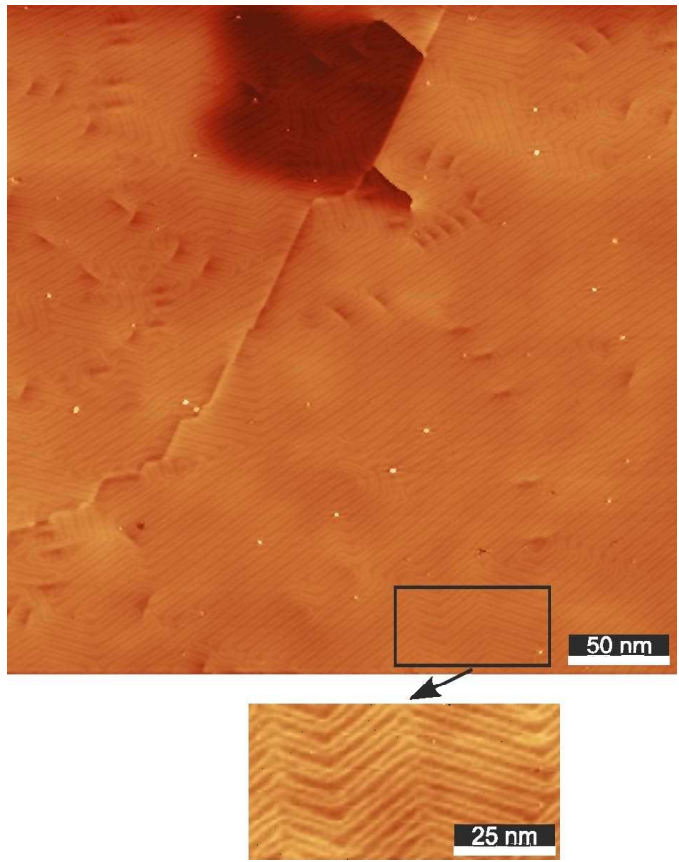}

\caption{\label{fig:STMAu} (Color online) \lengthnm{375x375} STM image of Au(111) grown on Mo(110) (sample voltage \unit[1]{V}, current \unit[0.25]{nA}). The gray scale has been adjusted between zero and the maximum height $Z_{max}$=1.5 $\AA$. A zoom of a selected area is show in the lower part, displaying the Au(111) herringbone reconstruction with $Z_{max}$=0.54 $\AA$.}
\end{figure}

Next a \thicknm{5}-thick Au film was deposited at RT on top of the Mo buffer layer.
The Au surface was then sputtered with \unit[1]{kV} Ar$^+$ ions to remove a few AL, then annealed at \tempdegC{550} during 30 minutes. The resulting Au surface was studied by
STM. It is atomically-flat with the usual $22\times\sqrt3$
reconstruction of Au\cite{bib-HEY1980,bib-SAN1991,bib-BAR1990}. However the presence of micro-grain-boundaries and dislocation loops prevent the occurrence of a perfect long range herringbone superstructure as for
Au single crystals (\figref{fig:STMAu}), its appearance being limited to restricted areas as shown in the zoom to \figref{fig:STMAu}.

\section{Co growth on Au(111)}

\begin{figure}
\includegraphics[width=3.4in]{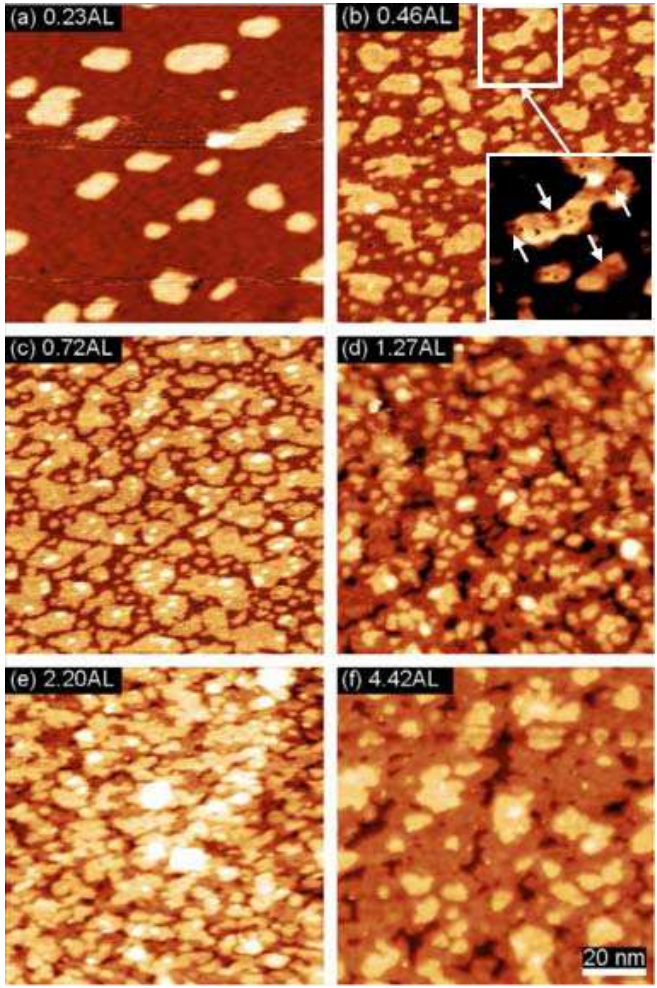}
\caption{\label{fig:Figcoau} (Color online) \lengthnm{100x100} STM images of PLD-grown Co/Au(111) films at various coverages, expressed in atomic layers (AL). After random nucleation at the first stages of growth (a), a bimodal distribution of islands with stripe-like corrugation is observed for submonoatomic coverings (b)-(c). The \lengthnm{25x25} zoom in (b) shows a stripe-like corrugation on the islands (note that the vertical scale was optimized in this case to better show the top of the islands. White arrows point at some of the lower areas on the islands). The growth of further layers proceeds close to a layer-by-layer fashion (d)-(f). The gray scale has been properly adjusted in each case between zero and the maximum
height $Z_{\mathrm{max}}$, with $Z_{\mathrm{max}}=4.2$, 5.3, 5.3, 5.5, 5.4 and $\unit[6.1]{\AA}$ for images (a), (b), (c), (d), (e) and (f) respectively. $Z_{\mathrm{max}}$ is $\unit[1.4]{\AA}$ for the inset to (b).}
\end{figure}

\figref{fig:Figcoau} shows STM pictures of various amounts of Co deposited at RT on the Au surfaces described above. In the sub-atomic-layer range Co growth proceeds through the more-or-less random nucleation of
1 AL-high islands. The distribution of island lateral size is quickly bimodal (\figref{fig:Figcoau} b-c) owing to the combination of homogeneous nucleation by adatom aggregation and heterogeneous nucleation on the Au(111) defects \cite{Nouvertne1999,Vardavas2004185}. The growth in this sub-atomic-layer range dramatically differs from the case of TD, for which the islands are 2 AL-high and nucleate almost solely at the elbows of the
Au herringbone reconstruction\cite{bib-VOI1991}, with a random 0.02 nm corrugation on the top of the islands owing to the large mismatch between Co and Au\cite{bib-REP2002b,bib-FRU2002}. Here, stripe-like areas of two different heights are observed on the top of the 1 AL-high Co islands, as the zoom of a selected area in \figref{fig:Figcoau} (b) shows (note that the vertical scale in this \lengthnm{25x25} area has been optimized to better display the top part of the islands). This feature points to some degree of intermixing between Co and Au as previously observed in PLD deposited systems\cite{bib-MEY2007}. As above mentioned, the interfacial intermixing in PLD is a result of the energy carried by the atoms or ions evaporated from the target and heated upon further interaction with the laser in the plume\cite{bib-SIN1990,bib-MEY2007}. Although we set the laser fluence just above the evaporation threshold, tails in the energy distribution or hot spots in the laser beam may create a small fraction of atoms or ions carrying a few eV or more. Notice that interfacial intermixing is not a systematic feature of PLD, as we did not observe such inhomogeneities in other systems, e.g. Fe\cite{bib-FRU2003d} or Co deposited on W
or Mo(110). The higher bonding energy of the latter with respect to Au may prevent the intermixing.
The growth mode of further Co layers proceeds close to a layer-by-layer fashion, with a morphology and inter-island distances at 4 AL of approximately 20 nm, very similar to those obtained with TD\cite{bib-FRU2003b}. Thus the topography of PLD-grown films a few AL thick is similar to their TD-grown counterparts, except for some intermixing at the bottom interface.

\section{Magnetic anisotropy and hysteresis}

\begin{figure}
  \begin{center}
    \includegraphics[width=3.2in]{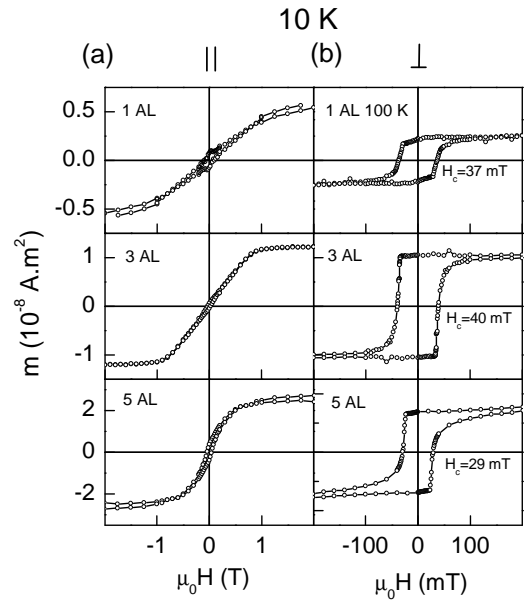}
    \caption{\label{fig:SQUID10K} SQUID hysteresis loops performed at low temperature for Co films
    of 1, 3 and \thickAL{5}, both in-plane (left column) and perpendicular to the plane (right column). All loops have been measured at \tempK{10} except
    the perpendicular loop of the \thickAL{1}-thick film, which was measured at \tempK{100}.}
  \end{center}
\end{figure}

\begin{figure}
  \begin{center}
    \includegraphics[width=3.2in]{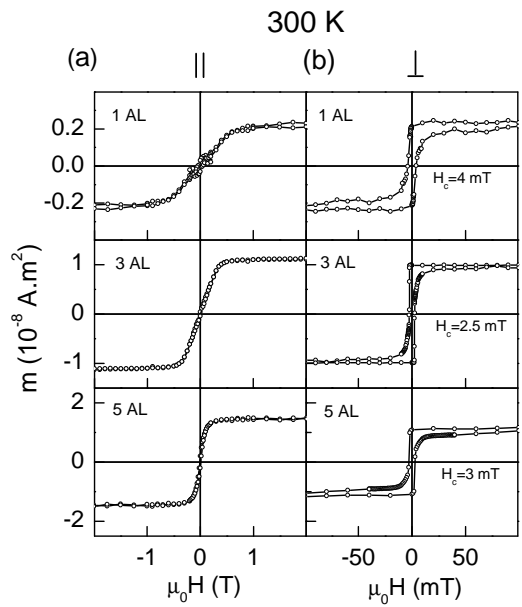}
    \caption{\label{fig:SQUID300K} SQUID hysteresis loops performed at RT for Co films
    of 1, 3 and \thickAL{5}, both in-plane (left column) and perpendicular to the plane (right column). }
  \end{center}
\end{figure}

Magnetization characterization was performed on samples similar to those presented above, however
immediately capped with a 2 nm thick Au layer \insitu for protection against contamination~(\ie, with no STM
investigation). The magnetization reversal of 1, 3 and 5 AL-thick Co samples was studied by means of SQUID magnetometry, under successively in-plane and perpendicular magnetic
fields, at low (\figref{fig:SQUID10K}) and RT (\figref{fig:SQUID300K}). Surprisingly, ferromagnetic behavior is found in all the PLD films at RT for Co coverings starting at 1 AL grown. This is an striking result since TD films of nominal thickness 1.5 AL and below are not ferromagnetic at RT\cite{bib-CHA1986}. The lower ferromagnetic critical thickness found in our PLD films stems from the different growth mode observed between both techniques. As above mentioned, for TD films Co nucleates in the form of isolated 2 AL-thick islands in relation to the herringbone reconstruction of Au(111)\cite{bib-HEY1980,bib-BAR1990}, forming crystallites of ultimately 6-10 nm in diameter\cite{bib-CES1989}, which remain superparamagnetic until percolation occurs around 1.6 ML\cite{bib-PAD1999b}. On the other hand, in our case PLD leads to 1 AL-high Co islands at the early stages of growth, with larger lateral extends as compared to TD films (irregularly shaped islands up to 25 nm long are found at coverings around 0.72 AL as shown in \figref{fig:Figcoau} (c)), with percolation starting at submonoatomic coverings, and thus favoring the observed ferromagnetic behavior. The accuracy of saturation magnetization $\Ms$ measurements using SQUID on magnetic thin films is subject to several factors, including error on the estimation of the volume of the magnetic films and background diamagnetic signals that must be subtracted. After correction of the measured data, saturation magnetization values close to the bulk value ($\Ms=\unit[1.446\times 10^{6}]{A/m}$) were found for the 3 and 5 AL-thick Co films, whereas a reduction around $\unit[30]{\%}$ was found for the 1 AL-thick film owing to the above mentioned interdiffusion of Co and Au, being this effect specially important in the low thickness regime.

Interestingly, the $\unit[100]{\%}$ remanence along the direction perpendicular to the plane and the almost closed loops in-the-plane demonstrate a full perpendicular anisotropy for all the coverings and measurement temperatures, as shown in \figref{fig:SQUID10K} and \figref{fig:SQUID300K}. We analyze the magnetic anisotropy energy (MAE) at low temperature to access the values in the fundamental state. The density of MAE was computed from the hysteresis loops along
the perpendicular direction as $K=\mu_{0}\int^{\Ms}_{0}H \diff{M}$.
The resulting values are summarized in \tabref{tab-anisotropy}. The MAE increases at low thickness,
consistently with the picture of a dominant contribution
of interface and/or magneto-elastic terms\cite{bib-GRA1968,bib-GRA1993}.

Concerning the easy axis of magnetization, the coercivity depends only weakly on the thickness. It decreases from around \unit[30-40]{mT} at low temperature to typically \unit[3-5]{mT} at RT. This sharp
decrease with temperature is common for ultrathin films because the activation volumes involved in nucleation or
activation processes underlying magnetization reversal\cite{bib-GIV2003} are small as they scale with the thickness of the film\cite{bib-FER1997,bib-BRU1990}. Thermal activation is therefore much enhanced compared to bulk materials. Previously Au/Co/Au films had been extensively prepared using TD\cite{bib-CHA1986} and more recently with ED\cite{bib-CAG2001}. Both exhibit perpendicular anisotropy up to about ten AL of Co. Their total MAE has been measured mostly for films 5 AL and thicker. For 5 AL figures in the range \unit[0.7-0.8]{MJ/m$^{3}$} have been given\cite{bib-CAG2001,bib-CHA1986}. This is roughly double the MAE of 5 AL-PLD-grown films (\tabref{tab-anisotropy}). This reduced value for our films may result from the intermixing at their lower interface. Comparison at lower thicknesses based on extrapolation using published
values of volume and surface energies may be hazardous and thus is not discussed here, although a similar reduction would be expected. The case of coercivity is more striking. The coercivity at RT of 3-5 AL TD or ED films are typically in the range 35-80 mT depending on
the preparation conditions\cite{bib-FER1997,bib-BAY1989,bib-FER1990}. This is more than one
order of magnitude higher than for our PLD films, an effect which obviously cannot be ascribed solely to the $\unit[50]{\%}$ decreased MAE. Instead we ascribe this to the difference
of growth modes, as it was done for the early onset of ferromagnetic behavior. In TD and ED the growth initially proceeds by the formation of 2 AL-high islands. This induces a significant roughness that may be responsible for the high coercivity, whereas the initial AL-growth with PLD yields smoother films, topographically and obviously magnetically. This allows the occurrence of ferromagnetism however still associated with a high coercivity as the domain wall propagation is hindered between islands. Already at 1 AL PLD films combine both ferromagnetism and a low coercivity owing to the initial growth mode in 1 AL-high islands.

Let us summarize the discussion of the relationship between the growth-modes-related microstructure and
magnetic properties. Owing to the intermixing-induced initial growth mode as 1 AL-islands, PLD films compared
to TD or ED films have ferromagnetic ordering already at 1 AL, a MAE reduced by $\approx$ $\unit[50]{\%}$ and a coercivity more than one order of magnitude lower.


\begin{table}
  \caption{Directly-measured total density of magnetic anisotropy energy at 10 K (in $\unit[]{MJ/m^3}$) for Au/Co/Au films fabricated by PLD.}
  \label{tab-anisotropy}
  \centering
  \begin{tabular}{ccc}
    \hline\hline
    \thickAL1 & \thickAL3 & \thickAL5 \\
    \hline
    0.91 &  0.65 &  0.32\\
    \hline\hline
  \end{tabular}
\end{table}

\section{Magneto-optical (MO) activity}

The magneto-optical (MO) properties of Au/Co/Au trilayers and Au/Co multilayers grown by TD as a function of the Co thickness have been extensively studied\cite{visnovsky:6783,Visnovsky1993179,Bib-Visnovsky1995,bib-Hamrle2001,bib-TAK1996,bib-CLA2008}. Vi\u{s}\u{n}ovsk\'y et al.\cite{visnovsky:6783,Visnovsky1993179,Bib-Visnovsky1995} showed that the MO response of these systems in the thin film limit is given by the sum of one independent plus one linearly dependent term on the Co film thickness. They ascribed the independent term to different Au/Co interface effects, such as Au-Co mismatch driven stress, lattice defects, Au-Co electronic orbital hybridization and intermixing at the interface\cite{visnovsky:6783}. More recently, Hamrle et al.\cite{bib-Hamrle2001} studied specifically the contribution of Co/Au interfaces to the MO response of Au/Co/Au trilayers, and concluded that the most important part of the interface contribution arises from intrinsic properties of the interface itself, i.e., from the Au-Co electronic hybridization. Nevertheless, no studies about the MO response of similar PLD grown structures have been reported so far.

\begin{figure}
\includegraphics[width=3.5 in]{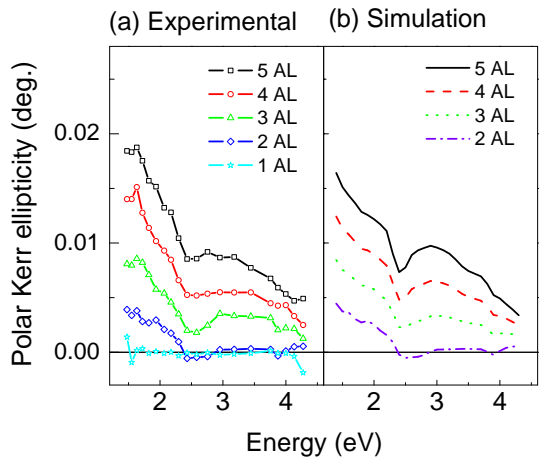}

\caption{\label{fig:MO} (Color online) (a) Experimental and (b) simulated polar Kerr ellipticity spectra for the systems with Co films ranging from 1 and \thickAL{5}.}
\end{figure}

Here, the MO characterization was performed on the samples presented in Section V, with graded Co thicknesses from 1 to 5 AL. Polar Kerr ellipticity spectra were measured in the spectral range from 1.4 to 4.3 eV. In this configuration, normal incident light is used, being the polarization change of the reflected light (rotation and ellipticity) measured with magnetic field applied perpendicular to the surface. As shown in \figref{fig:MO} (a), almost zero ellipticity is obtained for 1 AL, whereas a progressive increase is observed as the Co thickness increases over the entire spectral range, in accordance with previous reports on TD-grown systems\cite{bib-Hamrle2001,Bib-Visnovsky1995}. In addition, the characteristic peak arising from the Au plasma edge is found in all the cases around 2.5 eV\cite{bib-TAK1996,Bib-Visnovsky1995,visnovsky:6783,Visnovsky1993179}.

In order to model the evolution and shape of the ellipticity spectra,
simulations were performed using the transfer matrix formalism\cite{bib-SCH1996}. As shown in \figref{fig:MO} (a), a strong effect on the ellipticity spectra due to the Au/Co interfaces is observed for the trilayers with 1 and 2 AL thick Co films, exhibiting values smaller than expected for such Co thickness considering bulk MO constants\cite{Weller1994}, in agreement with previous reports\cite{visnovsky:6783}. In principal, the MO response of both Au/Co interfaces in the trilayers can be described by the measured MO response of the trilayer with a 2 AL thick Co film, since, as above mentioned, the MO response can be ascribed to short range Au-Co hybridization effects involving only the atomic layers closer to the interface, and in this case also some short range degree of intermixing as observed with STM. Thus, under the ultra thin film approach\cite{Bib-Visnovsky1995}, the ellipticity of the trilayers with 3, 4 and 5 AL layers thick Co films can be simulated by adding the ellipticity of 1, 2 and 3 Co thick continuous films respectively, addressing the inner part of the Co film unaffected by the interface, and the ellipticity of the trilayer with a 2 AL thick Co film, addressing the interface effects. The optical constants for Mo, Co and Au were obtained from Ref.\cite{bib-WEA1981} whereas bulk MO constants for Co were used\cite{Weller1994}. In fact, as shown in \figref{fig:MO} (b), simulations yield a good agreement with the measured spectra in intensity and shape for the simulated 3, 4 and 5 AL Co coverages. These results confirm that the MO response in the low Co thickness range is dominated by Au/Co interface effects, the ellipticity showing a linear response with the Co thickness for thicknesses starting at 3 AL Co, indicative of a continuous-film-like MO behavior in accordance with TD grown systems\cite{visnovsky:6783,bib-Hamrle2001,bib-TAK1996,Bib-Visnovsky1995}.

\section{Conclusion}

We fabricated epitaxial Co ultrathin films by pulsed-laser deposition on sapphire wafers buffered with a Au/Mo layer. Unlike TD-grown films which require 1.6 AL to become ferromagnetic, ferromagnetic behavior is found at room temperature for coverings starting at 1 AL. The films display perpendicular magnetization with magnetic anisotropy energy reduced by $\approx \unit[50]{\%}$ compared to TD-grown or electrodeposited films, and an unprecedented
low coercivity of $\approx \unit[5]{mT}$. We ascribed these differences to some degree of intermixing at the lower interface upon PLD. This reduces surface anisotropy, however promotes a layer-by-layer growth and thus yields a topographically and magnetically smoother film. The magneto-optical response in the low Co thickness range is dominated by Au/Co interface contributions. For thicknesses starting at 3 AL Co, the MO response has a linear dependence with the Co thickness, indicative of a continuous-film-like MO behavior.

\section{Acknowledgments}

We acknowledge financial support from FP6 EU-NSF program (STRP 016447 MagDot), French National
Research Agency (ANR-05-NANO-073 Vernanomag), ``FUNCOAT'' CONSOLIDER INGENIO 2010 CSD2008-00023, ``NANOMAGMA'' EU NMP-FP7-214107, ``MAGPLAS'' MAT2008-06765-C02-01/NAN, MAT2005-05524-C02-01 and ``NANOMAGNET'' CM S-0505/MAT/0194. CC acknowledges the Ministerio de Educaci\'{o}n y Ciencia through the FPI program and R. A. Lukaszew for financial support. We are grateful to J. Ferr\'{e} (LPS-Orsay) for both preliminary measurements and a critical reading of the manuscript.

\end{document}